# Analysis of half-spin particle motion in Reissner-Nordström and Schwarzschild fields by means of the method of effective potentials

M.V.Gorbatenko[1], V.P.Neznamov[1,2*], E.Yu.Popov[1]

[1]RFNC-VNIIEF, Russia, Sarov, Mira pr., 37, 607188
[2]National Research Nuclear University MEPhI, Moscow, Russia

Abstract

The paper presents the analysis of effective potentials of Dirac equations in Schwarzschild and Reissner-Nordström fields. It is shown that in the majority of the explored cases the condition of a particle "fall" to appropriate event horizons is fulfilled. The exception is one of the solutions with normalizable wave functions for the Reissner-Nordström extreme field, for which the existence of the stationary bound state of half-spin particles is possible inside the event horizon.

---

[*] E-mail: neznamov@vniief.ru

# Introduction

As it is well known, the form of static potentials in the Schrödinger non-relativistic equation qualitatively testifies to the presence or absence of bound states of particles, to the finite or infinite motion of particles, to the "fall" (localization) of the particle to a certain center, etc. (see, for instance, [1]).

The cases are known when the Dirac equation in a flat space was also analyzed in terms of the effective potential (EP). For instance, for the Coulomb potential in the Dirac equation the effective potentials were obtained and analyzed for atoms with nuclear charge $Z \geq 170$ (see the review [2], and the references therein). The necessary condition of such an analysis is the existence of the self-conjugate Hamiltonian of the Dirac equation and reality of the Dirac equation for real radial wave functions. The method of the effective potential can also be used in case of curved spaces. There are gravitational fields in which the analysis can be performed by means of EP method, among such fields are Schwarzschild and Reissner-Nordström (RN) gravitational fields. In these fields, as well as in Ref. [2], there are self-conjugate Dirac Hamiltonians allowing separation of angular and radial variables with differential equations with real coefficients for real radial wave functions.

Using of the EP method, allows us to obtain the following results. We show that in the majority of explored cases the condition of a particle "fall" to appropriate event horizons is fulfilled. The exception is one of the solutions with normalizable wave functions for the RN extreme field with the single event horizon. For this solution, inside the event horizon the possibility of existence of stationary bound states of half-spin particles is shown only at coinciding charge signs of the particle and the field source. At that, the region outside the event horizon is separated from the interior by an infinitely high potential barrier

The paper is organized as follows. In section 1 the RN metric, the self-conjugate Hamiltonian of a Dirac particle, the system of differential equations for radial wave functions are presented for reminding basic notions and facts. The features of the obtained effective potentials are revealed for different types of the RN field. The similar characteristics for the Schwarzschild field were obtained by the transition to the RN electrically uncharged field. In section 2 the obtained results are discussed.

# 1. Effective potentials for Reissner-Nordström and Schwarzschild fields

The Reissner-Nordström line element is

$$ds^2 = f_{R-N} dt^2 - \frac{dr^2}{f_{R-N}} - r^2\left(d\theta^2 + \sin^2\theta\, d\varphi^2\right), \tag{1}$$

where $f_{R-N} = \left(1 - \frac{r_0}{r} + \frac{r_Q^2}{r^2}\right)$, $r_0 = \frac{2GM}{c^2}$ is the gravitational radius of the Schwarzschild field, $r_Q = \frac{\sqrt{G}Q}{c^2}$ is the "charge" radius, $G$ is the gravitational constant, $c$ is the velocity of light.

1. If $r_0 > 2r_Q$, then

$$f_{R-N} = \left(1 - \frac{r_+}{r}\right)\left(1 - \frac{r_-}{r}\right), \tag{2}$$

where $r_\pm$ are inner and outer radii of the event horizons

$$r_\pm = \frac{r_0}{2} \pm \sqrt{\frac{r_0^2}{4} - r_Q^2}. \tag{3}$$

2. The case of $r_0 = 2r_Q$ corresponds to the RN extreme field with the radius of the event horizon $r_{extr} = r_0/2$.

3. The case of $r_0 < 2r_Q$ corresponds to the naked singularity without event horizons.

4. At $r_Q = 0$, the metric (1) is reduced to the Schwarzschild metric.

The Hamiltonian of the Schrödinger-type equation is self-conjugate. Consequently, the initial Dirac Hamiltonian in the RN field must be also self-conjugate for obtaining correct effective potentials.

The algorithms for obtaining self-conjugate Dirac Hamiltonians in external gravitational fields in which a pseudo-Hermitian quantum mechanics methods are employed are given in [3] - [5]. The self-conjugate Hamiltonian of massive and charged half-spin particle in the RN field was obtained in [6]:

$$H_\eta = H_\eta^+ = \sqrt{f_{R-N}}\,\beta m - i\alpha^1\left(f_{R-N}\frac{\partial}{\partial r} + \frac{1}{r} - \frac{r_0}{2r^2}\right) - \\ -i\sqrt{f_{R-N}}\frac{1}{r}\left[\alpha^2\left(\frac{\partial}{\partial \theta} + \frac{1}{2}\mathrm{ctg}\,\theta\right) + \alpha^3\frac{1}{\sin\theta}\frac{\partial}{\partial \varphi}\right] + \frac{eQ}{r}. \tag{4}$$

In (4) $\alpha^k, \beta$ are Dirac matrices. After separation of variables we obtain system of first order differential equations for radial functions $F_{R-N}(\rho), G_{R-N}(\rho)$

$$f_{R-N}\frac{dF_{R-N}(\rho)}{d\rho}+\left(\frac{1+\kappa\sqrt{f_{R-N}}}{\rho}-\frac{\alpha}{\rho^2}\right)F_{R-N}(\rho)-\left(\varepsilon-\frac{\alpha_{em}}{\rho}+\sqrt{f_{R-N}}\right)G_{R-N}(\rho)=0,$$

$$f_{R-N}\frac{dG_{R-N}(\rho)}{d\rho}+\left(\frac{1-\kappa\sqrt{f_{R-N}}}{\rho}-\frac{\alpha}{\rho^2}\right)G_{R-N}(\rho)+\left(\varepsilon-\frac{\alpha_{em}}{\rho}-\sqrt{f_{R-N}}\right)F_{R-N}(\rho)=0,$$

(5)

in (5), dimensionless variables are:

$$\rho=\frac{r}{l_c};\ \varepsilon=\frac{E}{m};\ \alpha=\frac{r_0}{2l_c}=\frac{GMm}{\hbar c}=\frac{Mm}{M_P^2};$$

(6)

$$\alpha_Q=\frac{r_Q}{l_c}=\frac{\sqrt{G}Qm}{\hbar c}=\frac{\sqrt{\alpha_{fs}}}{M_P}m\frac{Q}{|e|};\ \alpha_{em}=\frac{eQ}{\hbar c}=\alpha_{fs}\frac{Q}{e},$$

(7)

$l_c=\frac{\hbar}{mc}$ is the Compton wave-length of the Dirac particle; $E$ is the energy of the Dirac particle;

$M_P=\sqrt{\frac{\hbar c}{G}}$ is the Planck mass; $\alpha_{fs}\approx 1/137$ is the electromagnetic fine structure constant;

$\alpha, \alpha_{em}$ are gravitational and electromagnetic coupling constants; $\alpha_Q$ is the dimensionless constant characterizing the electromagnetic field source in the RN metric.

The separation constant $\kappa$ is

$$\kappa=\mp\left(j+\frac{1}{2}\right)=\begin{cases}-(l+1),& j=l+\frac{1}{2}\\ l,& j=l-\frac{1}{2}\end{cases};$$

$l, j$ are quantum numbers of orbital and total angular momentums of the Dirac particle.
We will use the system of units $\hbar=c=1$.

Let us write

$$f_{R-N}=1-\frac{2\alpha}{\rho}+\frac{\alpha_Q^2}{\rho^2}=\left(1-\frac{\rho_+}{\rho}\right)\left(1-\frac{\rho_-}{\rho}\right),$$

(8)

where

$$\rho_+=\alpha+\sqrt{\alpha^2-\alpha_Q^2}\quad\text{is the radius of the outer event horizon,}$$

(9)

$$\rho_-=\alpha-\sqrt{\alpha^2-\alpha_Q^2}\quad\text{is the radius of the inner event horizon.}$$

(10)

Rewrite the system of the first order differential equations (5) as of the second order differential equation and then using substitution

$$\psi(\rho)=F(\rho)\exp\left(\frac{1}{2}\int A_1(\rho')d\rho'\right),$$

(11)

we obtain the Schrödinger-type equation

$$\frac{d^2\psi(\rho)}{d\rho^2}+2\left(E_{schr}-U_{eff}(\rho)\right)\psi(\rho)=0,$$

(12)

where

$$E_{schr} = \frac{1}{2}(\varepsilon^2 - 1),$$

$$U_{eff}(\rho) = E_{schr} + \frac{1}{4}\frac{dA_1(\rho)}{d\rho} + \frac{1}{8}A_1^2(\rho) - \frac{1}{2}B_1(\rho). \quad (13)$$

In Eqs. (11) and (13)

$$A_1(\rho) = -\frac{1}{B(\rho)}\frac{dB(\rho)}{d\rho} - A(\rho) - D(\rho),$$

$$B_1(\rho) = -B(\rho)\frac{d}{d\rho}\left(\frac{A(\rho)}{B(\rho)}\right) - C(\rho)B(\rho) + A(\rho)D(\rho). \quad (14)$$

In Eq. (14)

$$A(\rho) = -\frac{1}{f_{R-N}}\left(\frac{1 + \kappa\sqrt{f_{R-N}}}{\rho} - \frac{\alpha}{\rho^2}\right),$$

$$B(\rho) = \frac{1}{f_{R-N}}\left(\varepsilon - \frac{\alpha_{em}}{\rho} + \sqrt{f_{R-N}}\right),$$

$$C(\rho) = -\frac{1}{f_{R-N}}\left(\varepsilon - \frac{\alpha_{em}}{\rho} - \sqrt{f_{R-N}}\right), \quad (15)$$

$$D(\rho) = -\frac{1}{f_{R-N}}\left(\frac{1 - \kappa\sqrt{f_{R-N}}}{\rho} - \frac{\alpha}{\rho^2}\right).$$

Eqs. (13) - (15) show that the effective potentials depend on $\rho$ and on five parameters $\kappa, \alpha, \alpha_Q, \alpha_{em}, \varepsilon$. Let us note that the energy of the Dirac particle $\varepsilon = \frac{E}{m}$ is also a parameter, $\varepsilon < 1$ corresponds to the bound states of a particle. Let us explore the behavior of the effective potentials in the vicinity of their singularities.

## 1.1 The Reissner-Nordström field with two event horizons ($\alpha^2 > \alpha_Q^2$)

The effective potentials have the following leading terms as $\rho \to \rho_-$ ($\rho < \rho_-$) and $\rho \to \rho_+$ ($\rho > \rho_+$)

$$U_{eff}^{R-N}(\rho)\big|_{\rho \to \rho_-} = -\frac{1}{(\rho - \rho_-)^2}\left[\frac{1}{8} + \frac{1}{8}\frac{(\varepsilon\rho_- - \alpha_{em})^2 \rho_-^2}{(\alpha - \rho_-)^2}\right] + O\left(\frac{1}{\rho - \rho_-}\right), \quad (16)$$

$$U_{eff}^{R-N}(\rho)\big|_{\rho \to \rho_+} = -\frac{1}{(\rho_+ - \rho)^2}\left[\frac{1}{8} + \frac{1}{8}\frac{(\varepsilon\rho_+ - \alpha_{em})^2 \rho_+^2}{(\rho_+ - \alpha)^2}\right] + O\left(\frac{1}{\rho_+ - \rho}\right). \quad (17)$$

The numerators in the Eqs. (16) and (17) are always $\geq \frac{1}{8}$ at any values $\alpha, \alpha_Q, \alpha_{em}$ and $\varepsilon$.

It is well known [1] that quantum-mechanical particle "falls" to the center if the singular effective potential behaves as $\left(-C/\rho^2\right)$ with $C > 1/8$, otherwise it does not "fall" to center. The above expressions for singular effective potentials behave as $\left(-C_1/(\rho-\rho_\pm)^2\right)$ in the vicinities of horizons. If in (16), (17) $\varepsilon \neq \dfrac{\alpha_{em}}{\rho_-}$ and $\varepsilon \neq \dfrac{\alpha_{em}}{\rho_+}$, then $C_1 > 1/8$. It means that Dirac's particles "fall" to the horizons (either inner or outer). "Fall" to the horizon in the RN background is avoided if $\varepsilon = \dfrac{\alpha_{em}}{\rho_-}$ in the interior region or $\varepsilon = \dfrac{\alpha_{em}}{\rho_+}$ in the exterior region. However, the solutions with these energy levels fails to be normalizable [8].

For the Schwarzschild field, $\rho_+ \equiv \rho_0 = 2\alpha$, $\rho_- = 0$, $\alpha_{em} = 0$. In the region $\rho > 2\alpha$ the effective potential (17) in the vicinity of event horizon has a form

$$U_{eff}^{R-N}(\rho)\big|_{\rho \to 2\alpha} = -\frac{1}{(\rho-2\alpha)^2}\left[1/8 + 2\alpha^2\varepsilon^2\right] + O\left(\frac{1}{\rho-2\alpha}\right). \quad (18)$$

In this case, the particle will always be in the mode of a "fall" to the event horizon with $\rho_0 = 2\alpha$.

For the Schwarzschild field, at $\rho < 2\alpha$ and for the RN field at $\rho_- < \rho < \rho_+$, the effective potentials become complex. In these cases, the standard analysis of the motion of a quantum-mechanical particle by means of the effective potentials of the Dirac equation seems impossible. Let us note that the region of $\rho < 2\alpha$ for the Schwarzschild field and the region of $\rho_- < \rho < \rho_+$ for the RN field correspond to the values $g_{00} < 0$.

### 1.2 The Reissner-Nordström extreme field ($\alpha^2 = \alpha_Q^2$)

In this case, there is the single event horizon with the radius $\rho_+ = \rho_- = \alpha$.

If $\varepsilon \neq \dfrac{\alpha_{em}}{\alpha}$ and $\rho \to \alpha$, the effective potential is

$$U_{eff}^{R-N}(\rho)\big|_{\rho \to \alpha} = -\frac{\left(\varepsilon - \dfrac{\alpha_{em}}{\alpha}\right)^2 \alpha^4}{2(\rho-\alpha)^4} - \frac{\left(\varepsilon - \dfrac{\alpha_{em}}{\alpha}\right)\alpha^3\left(2\varepsilon - \dfrac{\alpha_{em}}{\alpha}\right)}{(\rho-\alpha)^3} + O\left(\frac{1}{(\rho-\alpha)^2}\right). \quad (19)$$

The effective potential in the vicinity of horizon is more singular than $\left(-L/(\rho-\alpha)^2\right)$ and it means that Dirac's particle "falls" to horizon. It is easy to see from (19) that if we take into account solution [8,9] $\varepsilon = \dfrac{\alpha_{em}}{\alpha}$ then the first and second terms are vanishing, and expression for the effective potential becomes

$$U_{\text{eff}}^{R-N}(\rho) = \frac{1}{2}\left[\frac{\left(1-\dfrac{\alpha_{em}^2}{\alpha^2}\right)\rho^4 + \left(\kappa^2+\kappa\right)\rho^2 - \alpha(\kappa+1)\rho + \dfrac{3}{4}\alpha^2}{\rho^2(\rho-\alpha)^2} - \left(1-\dfrac{\alpha_{em}^2}{\alpha^2}\right)\right]. \qquad (20)$$

The leading term of the potential (20) as $\rho \to \alpha$ becomes

$$\left. U_{\text{eff}}^{R-N}(\rho)\right|_{\rho \to \alpha} = -\frac{\frac{1}{4}-\kappa^2-\alpha^2+\alpha_{em}^2}{2(\rho-\alpha)^2}. \qquad (21)$$

According to conventional quantum-mechanical analysis, see e.g. [1], "fall" to the center (to the horizon in the present case) is prevented provided that condition

$$\kappa^2 + \alpha^2 - \alpha_{em}^2 > 0. \qquad (22)$$

However the eigenvalue $\varepsilon = \dfrac{\alpha_{em}}{\alpha}$ with the convergent normalization integral for the wave function to exist, a more strong condition [8] should be imposed:

$$\kappa^2 + \alpha^2 - \alpha_{em}^2 > \frac{1}{4}. \qquad (23)$$

On the other hand, we claim that in the exterior region of an extreme RN metric a stable bound states described by the Dirac equation do not exist. This is consistent with the results of Refs. [7] and [9]. Indeed, let us consider two possibilities. First, $\alpha > |\alpha_{em}|$. In this case, the found effective potential has no extremal point. The effective potential $U_{\text{eff}}^{R-N}(\rho)$ is schematically shown in figure 1.

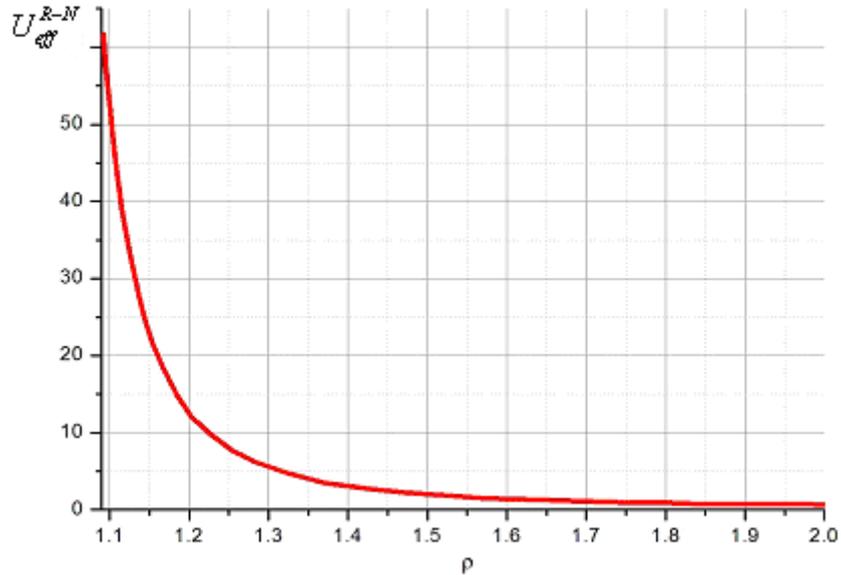

Figure 1. Behavior of the effective potential of the Schrödinger-type equation in the exterior of the RN extreme field as $\alpha > \alpha_{em}$, $\alpha = \alpha_Q = 1$, $\alpha_{em} = 0.9$, $\varepsilon = 0.9$.

It is seen from this plot for the effective potential that there are no stationary bound states in the exterior region of extreme RN metric.

Second possibility is $\alpha < |\alpha_{em}|$. In this case the shape of the potential $U_{eff}^{R-N}(\rho)$ varies qualitatively (see figure 2).

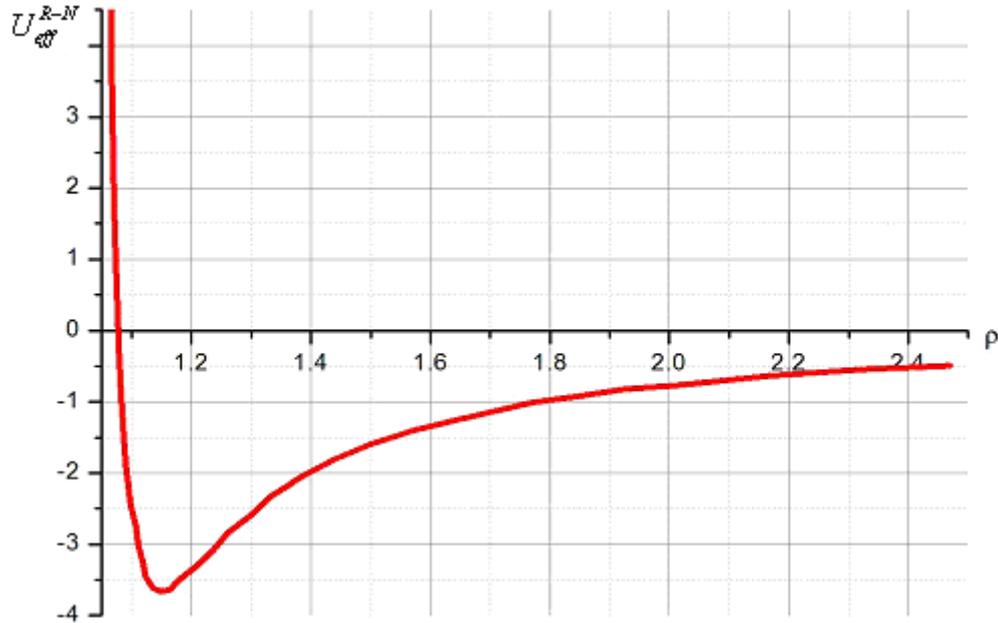

Figure 2. Behavior of the effective potential of the Schrödinger-type equation in the field of the RN extreme field as $\alpha < \alpha_{em}$, $\alpha = \alpha_Q = 1$, $\alpha_{em} = 1.25$, $\varepsilon = 1.25$.

Can the stable energy level $\varepsilon = \dfrac{\alpha_{em}}{\alpha}$ exist in the case $\alpha < |\alpha_{em}|$? Let us first assume that the field source and the particle have like signs of electric charges, $\alpha_{em} > 0$. Clearly, this energy corresponding to a bound states can not exist because $\varepsilon > 1$ so that this energy level belongs to the continuum part of the spectrum. Let the field source and the particle have opposite signs of electric charges, $\alpha_{em} < 0$. It follows that $\varepsilon = -\left|\dfrac{\alpha_{em}}{\alpha}\right| < -1$, which formally does not prevent to the existence a discrete energy level with $E < -m$. For a Dirac particle, outside the event horizon, at $\varepsilon = -\dfrac{|\alpha_{em}|}{\alpha} < -1$ ($E < -m$) the vacuum reaction similar to that of electron-positron vacuum in atoms with super-heavy nuclei with $Z \geq 170 \div 175$ [11] - [14].

Let us consider the interior region of an extreme Reissner-Nordsröm metric ($0 < \alpha < \rho$). The effective potential is positive (see figure 3).

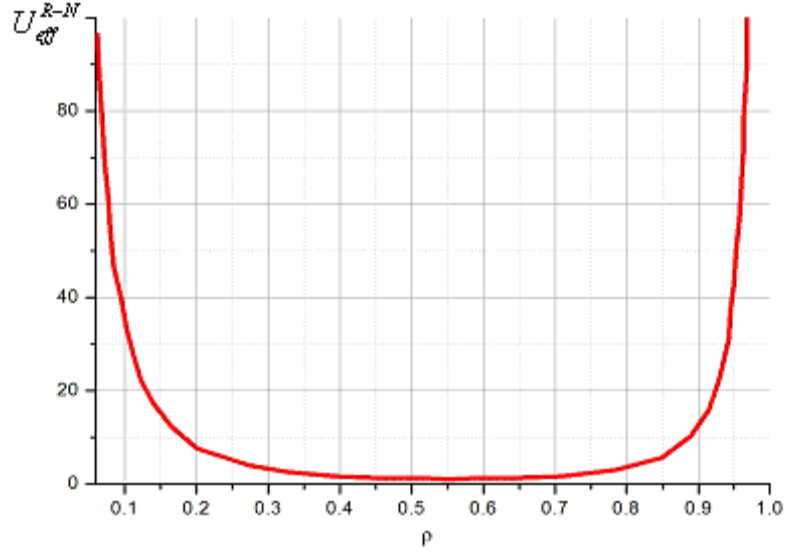

Figure 3. Behavior of the effective potential of the Dirac equation in the RN extreme field inside the event horizon $0 < \rho < \alpha$, $\alpha = \alpha_Q = 1$, $\alpha_{em} = 1.25$, $\varepsilon = 1.25$.

In the vicinity of the origin, we have $U_{eff}^{R-N}(\rho)\big|_{\rho \to 0} = \frac{3}{8}\frac{1}{\rho^2} + O\left(\frac{1}{\rho}\right)$ while in the vicinity of the horizon $U_{eff}^{R-N} \sim 1/(\rho-\alpha)^2$. Can a discrete state with $\varepsilon = \frac{\alpha_{em}}{\alpha}$ exist there? The answer is positive provided that the particle and the source of the field have the same sign of their electric charges. Otherwise the answer is negative. Note that the energy level must be greater than the minimum of the effective potential, that is

$$\frac{\alpha_{em}}{\alpha} > \min U_{eff}^{R-N}(\rho), \ 0 \leq \rho < \alpha. \tag{24}$$

## 2. Discussions

The properties of solutions to the Dirac equation in static gravitational fields described by Schwarzschild and Reissner-Nordström metrics were analyzed by using the method of the effective potentials. The performed analysis allowed us to make the following conclusions. For the majority of explored metrics, allowing event horizons, the motion of quantum-mechanical particles is implemented in the mode of a "fall" to the appropriate event horizons. The absence of the mode of the particle "fall" to the event horizons is implemented in the RN field only in two cases. The first case corresponds to the solutions with eigenvalues $\varepsilon = \frac{\alpha_{em}}{\rho_-}$ and $\varepsilon = \frac{\alpha_{em}}{\rho_+}$ for the RN field with the event horizons $\rho_+, \rho_-$ at $\alpha^2 > \alpha_Q^2$. However, these solutions are irregular because of the divergence of the normalization integrals of wave functions. Another possibility for avoiding the "fall" to the horizon in RN field is the solution $\varepsilon = \dfrac{\alpha_{em}}{\alpha}$, which corresponds to the case of extreme RN metric ($\alpha^2 = \alpha_Q^2$) when two horizons merge, $\rho_+ = \rho_- = \alpha$. We arrive at the conclusion that bound states do not exist outside the event horizon, but a single state can exist in the interior region of the extreme RN metric. If $\kappa^2 + \alpha^2 - \alpha_{em}^2 > 1/4$ and $\varepsilon = \dfrac{\alpha_{em}}{\alpha}$, then the normalization integral is convergent and the wave functions turn into zero at event horizon. The effective potential as $\rho = \alpha$ exhibits the second order pole with a positive coefficient. Because of the zero probability densities at event horizon this potential barrier becomes quantum-mechanically impenetrable [15].

Realization of the RN extreme field is connected with certain exotics. In compliance with (6), (7), the condition $\alpha^2 = \alpha_Q^2$ leads to the mass of the gravitational-field source

$$M = \sqrt{\alpha_{fs}} M_P \left|\dfrac{Q}{e}\right|.$$

It is seen that the mass $M$ must be comparable with or more than the Planck mass.

The ratio of $\dfrac{\alpha_{em}}{\alpha}$ can be expressed as $\dfrac{\alpha_{em}}{\alpha} = \sqrt{\alpha_{fs}} \dfrac{M_P}{m} \operatorname{sgn} \dfrac{Q}{e}$. For charged particles of the Standard Model, this relation is very large. For instance, for an electron

$$\dfrac{\alpha_{em}}{\alpha} = \dfrac{1.2 \cdot 10^{22}}{11.7 \cdot 0.5} \operatorname{sgn} \dfrac{Q}{e} \approx 2 \cdot 10^{21} \operatorname{sgn} \dfrac{Q}{e},$$

it follows that stationary level $\varepsilon = \dfrac{\alpha_{em}}{\alpha}$ in the interior region of extreme RN metric has extremely high magnitude $E = \varepsilon m \approx 2 \cdot 10^{21} m$. The case $\dfrac{|\alpha_{em}|}{\alpha} < 1$ considered in item 1.2 can be fulfilled only if a particle with an elementary charge $e$ and a mass comparable with the Plank mass $M_P$ is used as the Dirac particle.

The structure with the bound states of Dirac particles in the extreme RN background was proposed as a candidate for the dark matter [10]. The drawback of this proposal is non-compensation of the particle charges and of the RN electrical-field source which contradicts the observations of electrical neutrality of the dark matter.

## **Acknowledgements**

The authors would like to thank Prof. P. Fiziev for the incentive discussions at the initial stage of the paper-related efforts and A.L. Novoselova for the essential technical support while elaborating the paper.